\documentclass[final,a4paper,3p]{elsarticle}

\journal{arxiv.org}

\usepackage{array}                   % customize tables and arrays
\usepackage{amsmath,amssymb,amsfonts,amsthm}
\allowdisplaybreaks
\usepackage{dsfont,bm,bbm} 
\usepackage{relsize}
\usepackage{mathrsfs}
\usepackage{booktabs}
\usepackage{mathtools}
\usepackage[german,english]{babel}   % language to use in the document
\usepackage{caption}                 
\usepackage[utf8]{inputenc}          % special characters and encoding
\usepackage[T1]{fontenc}
\usepackage{float}
\usepackage{etoolbox}
\usepackage{graphicx}                % graphics
\usepackage{lmodern}                 % some fonts
\usepackage{microtype}    % controls spaces between words and paragraphs
\usepackage{lineno}
\usepackage{multirow}
\usepackage{calligra}
\usepackage{nicefrac}
\usepackage[usenames,dvipsnames]{xcolor}
\usepackage{tikz}
\usepackage{tikz-network}
\usepackage{algorithm}
\usepackage[noend]{algpseudocode}
\usepackage{subcaption}

\usepackage[deletedmarkup=xout,authormarkup=none]{changes}
\usepackage{url} 
\usepackage[colorlinks]{hyperref}
\usepackage{cleveref}
\crefformat{equation}{(#2#1#3)}
\AtBeginDocument{%
  \hypersetup{
    citecolor=Black,
    linkcolor=Black,   
    urlcolor=Violet,
	pdfauthor={Uribe et al.}}}


\makeatletter
\def\url@leostyle{%
\@ifundefined{selectfont}{\def\UrlFont{\sf}}{\def\UrlFont{\small\ttfamily}}}
\makeatother
\newcommand*\patchAmsMathEnvironmentForLineno[1]{%
  \expandafter\let\csname old#1\expandafter\endcsname\csname #1\endcsname
  \expandafter\let\csname oldend#1\expandafter\endcsname\csname end#1\endcsname
  \renewenvironment{#1}%
     {\linenomath\csname old#1\endcsname}%
     {\csname oldend#1\endcsname\endlinenomath}}% 
\newcommand*\patchBothAmsMathEnvironmentsForLineno[1]{%
  \patchAmsMathEnvironmentForLineno{#1}%
  \patchAmsMathEnvironmentForLineno{#1*}}%
\AtBeginDocument{%
\patchBothAmsMathEnvironmentsForLineno{equation}%
\patchBothAmsMathEnvironmentsForLineno{align}%
\patchBothAmsMathEnvironmentsForLineno{flalign}%
\patchBothAmsMathEnvironmentsForLineno{alignat}%
\patchBothAmsMathEnvironmentsForLineno{gather}%
\patchBothAmsMathEnvironmentsForLineno{multline}%
}

\begin{document}

\begin{frontmatter}

\title{Identifiability and amortized inference limitations in Kuramoto models}

\author[label1,label2]{Emma Hannula\corref{cor1}}
\ead{emma.hannula@lut.fi}
\cortext[cor1]{Corresponding author.}

\author[label1,label2]{Jana de Wiljes}
\author[label1]{Matthew T. Moores}
\author[label1]{Heikki Haario}
\author[label1,label3]{Lassi Roininen}

\affiliation[label1]{organization={LUT University},
            addressline={Yliopistonkatu 34},
            city={Lappeenranta},
            postcode={53850},
            country={Finland}}

\affiliation[label2]{organization={TU Ilmenau},
            addressline={Ehrenbergstraße 29},
            city={Ilmenau},
            postcode={98693},
            country={Germany}}

\affiliation[label3]{organization={University of Cambridge},
            addressline={Herschel Road},
            city={Cambridge},
            postcode={CB3 9AL},
            country={United Kingdom},}

\begin{abstract}
Bayesian inference is a powerful tool for parameter estimation and uncertainty quantification in dynamical systems. However, for nonlinear oscillator networks such as Kuramoto models, widely used to study synchronization phenomena in physics, biology, and engineering, inference is often computationally prohibitive due to high-dimensional state spaces and intractable likelihood functions. We present an amortized Bayesian inference approach that learns a neural approximation of the posterior from simulated phase dynamics, enabling fast, scalable inference without repeated sampling or optimization. Applied to synthetic Kuramoto networks, the method shows promising results in approximating posterior distributions and capturing uncertainty, with computational savings compared to traditional Bayesian techniques. These findings suggest that amortized inference is a practical and flexible framework for uncertainty-aware analysis of oscillator networks.
\end{abstract}

\end{frontmatter}

\section{Introduction}

Bayesian inference uses probability theory to update beliefs about unknown parameters $\theta$ based on observed data $y$. It is based on Bayes' theorem that combines prior distribution $p(\theta)$ and the likelihood function $p(y \mid \theta)$ to produce a posterior distribution. Bayes' theorem is defined as:

\begin{equation}\label{eq:posterior}
    p(\theta \mid y) = \frac{p(y \mid \theta)p(\theta) }{p(y)},
\end{equation}
where $p(y) = \int p(y \mid \theta)p(\theta) d\theta$ is the marginal likelihood that acts as a normalizing constant \cite{Gelman2013bayesian}. 

Traditional Bayesian inference aims to estimate the posterior distribution separately for each new data instance. Because posterior inference requires either Monte-Carlo methods or other complicated algorithms, the cost of the simulation increases with respect to the new data \cite{Gelman2013bayesian, Li2021MCMCreview}.

\begin{figure}
    \centering
    \includegraphics[width=\linewidth]{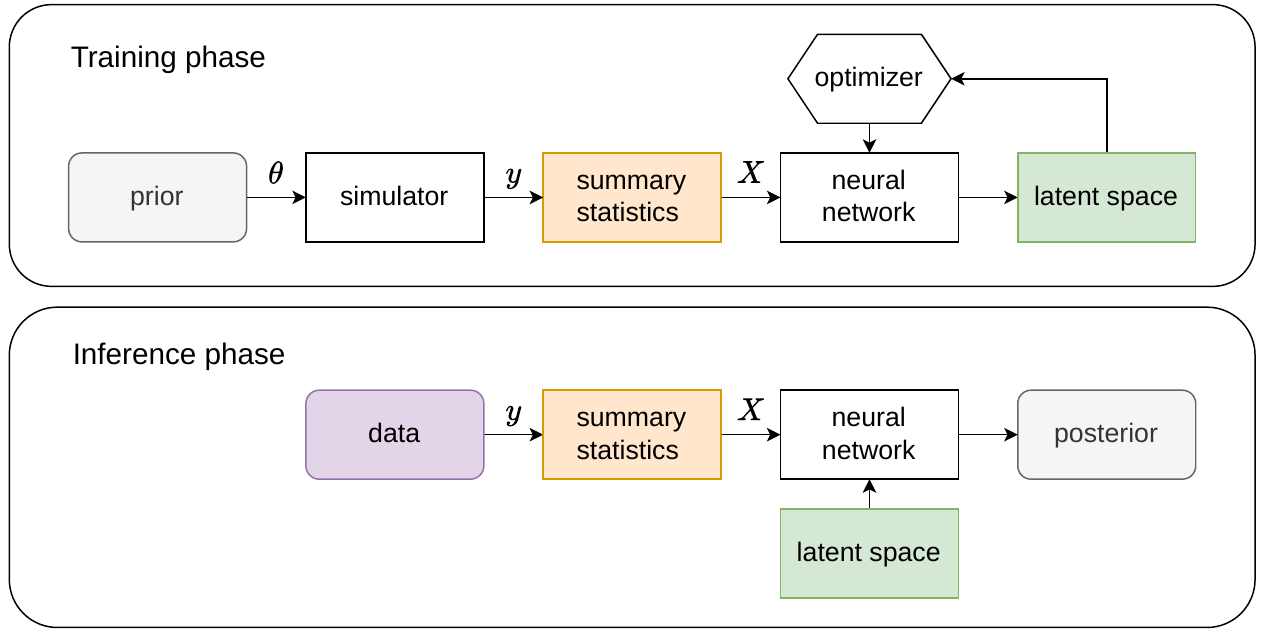}
    \caption{Illustration of training and inference phases in amortized Bayesian inference.}
    \label{fig: abi illustration}
\end{figure}

A promising approach to addressing the problem of complex and intractable probability distributions is by using machine learning methods, such as neural networks \cite{zammit2025neural}. In this work, neural networks are used as surrogate models for the probability densities required for likelihood evaluation, thereby avoiding the need to compute high-dimensional integrals explicitly. These networks are typically trained on data generated from the underlying statistical models, a setting commonly called simulation-based inference (SBI) \cite{Cranmer_2020}. After training, the neural surrogates can efficiently approximate quantities that would otherwise require expensive re-computation for each new dataset. This approach moves the computationally expensive inference to the training phase, resulting in more efficient inference time. This principle is known as amortization and is the basis of amortized Bayesian inference (ABI). The illustration of the phases in this work is visualized in Figure \ref{fig: abi illustration}.

Recent developments in amortized Bayesian inference have enabled efficient and flexible inference for a range of complex dynamical systems \cite{Rahmandad2025incorporating}. However, explicit applications to known synchronization models, for example Kuramoto and Lorenz systems, are still limited. 
%nonlinear models, for example the Kuramoto, Lorenz, and Lotka–Volterra systems, are still limited.

In this article, we apply amortized Bayesian inference to Kuramoto networks, introduced by Kuramoto et al.\ \cite{Kuramoto_1975}, to infer posterior distributions over model parameters from observed phase dynamics. Our approach leverages modern neural posterior estimation techniques trained on simulated data, enabling efficient inference while retaining a principled Bayesian treatment of uncertainty. 

\paragraph{Related work}

Many simulation-based inference methods have been developed in recent years to approximate either the likelihood function \cite{Papamakarios2019aNLE, Hermans2020} or the posterior distribution directly \cite{Orozco_2025, Papamakarios2016NPE}. In addition, many modern approaches employ amortized inference to enable efficient reuse of learned inference models across datasets \cite{Papamakarios2019aNLE, bayesflow_2020_original, radev2020amortized}. To overcome the problem of dealing with missing or unstructured data in ABI, Gloeckler et al.\ \cite{Gloeckler_2024} introduced all-in-one SBI, where they use a combination of transformers and probabilistic diffusion models. 

ABI has previously been applied to ecological models (Lotka-Volterra) \cite{Gloeckler_2024}, epidemiological models (SIRD, SEIRb) \cite{Gloeckler_2024, Rahmandad2025incorporating} and models that arise in the field of medicine, including cognitive science \cite{Orozco_2025, radev2020amortized}.

\subsection{Contributions}

In this work, we propose amortized Bayesian inference for parameter estimation of a Kuramoto network of oscillators. The main contributions are as follows: 
\begin{itemize}
    \item Implementation of amortized Bayesian inference for Kuramoto network parameter estimation.
    \item Model extension of parameter estimation for complex systems.
    \item Comparison of the methods’ ability to recover the parameters is performed against the MCMC approach \cite{Shah2023kuramoto,shah2023fods}.
    \item Feature engineering for summary statistics and comparison with summary networks.
\end{itemize}
The remainder of the paper is structured as follows. In Section \ref{sec: Kuramoto}, we introduce different aspects of the Kuramoto model. In Section \ref{sec: Method}, we implement our ABI approach to parameter estimation of a Kuramoto network. Section \ref{sec: Experiments} validates the approach with our experiments. Finally, in Section \ref{sec: Conclusions} we summarize the main
conclusions and results of the paper and discuss potential avenues for future work.

\section{Kuramoto network of oscillators}
\label{sec: Kuramoto}

The Kuramoto model is a mathematical framework used to study synchronization in systems of coupled oscillators. It is widely applied across fields such as physics, biology, engineering, and the social sciences because it explains how individual elements in a system can spontaneously align their behavior \cite{Wu2020Kuramoto, Nicosia2013}.

A first order network consisting of Kuramoto trajectories from $N$ oscillators can be described by the differential equation \cite{Kuramoto_1975}:
\begin{equation}
\frac{d \psi_i}{dt} = \omega_i + \frac{\kappa}{N} 
\sum_{j=1}^{N}\sin(\psi_j-\psi_i), \quad i = 1,\dots,N,
\label{eq: Kuramoto1}
\end{equation}
where $\omega_i$ denotes the natural frequency of oscillator $i$, $\psi_i$ is the phase and $\kappa$ is the coupling constant. 
The coupling constant defines how strongly the oscillators are coupled to each other. 
Later, Kuramoto introduced order parameter $r$, where the relative $\kappa$ strength is determined through the oscillators' phase $\psi_i$ and the mean phase $\Psi$ \cite{Kuramoto_2003}:

\begin{equation}
    r\exp(i\Psi) = \frac{1}{N}\sum_{j=1}^{N} \exp(i\psi_j).
\end{equation}

The Kuramoto model can be transformed into the mean-field transformed Kuramoto. Multiplying by $\exp(i\psi_j)$ and equating the imaginary part, the Equation \eqref{eq: Kuramoto1} can be rewritten in an angular phase displacement equation
\begin{equation}
\frac{d \psi_i}{dt} = \omega_i + \kappa r \sin(\Psi-\psi_i), \quad i = 1,\dots,N.
\label{eq: Kuramoto2}
\end{equation}
The mean-field transformed Kuramoto is numerically faster to calculate than the original form in Equation \eqref{eq: Kuramoto1} \cite{Shah2023kuramoto}. 

\begin{figure}[!ht]
    \centering
    \includegraphics[width=\linewidth]{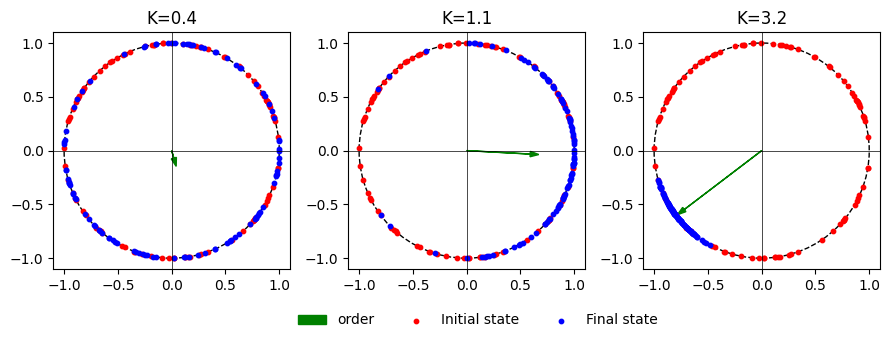}
    \caption{Three simulations from the Kuramoto system  with $N=100$ oscillators with given initial values and frequencies, randomized by  $\zeta \sim \mathcal{N}(0,10^{-2})$. The system is simulated for the duration of 1000 time steps.}
    \label{fig: Kuramoto}
\end{figure}

To study the impact of the synchronization parameter $\kappa$, we randomize the initial values  $ \psi_i(0) \sim \mathcal{N}(0, \sigma^2)$. For large $N$, the long-term statistical behavior is largely independent of initial conditions. The rotation frequencies $\omega_i$ can be fixed or randomized, too.  
The noisy first order Kuramoto system is visualized in Figure \ref{fig: Kuramoto}. It shows three different first and final states of 100 oscillators with varying coupling parameters, illustrating its effect in the final state.

\paragraph{Complex network of Kuramoto oscillators}

In the simple first order Kuramoto system, all of the oscillators have the same coupling parameter $\kappa$. If we wish to simulate a more complex system, where the oscillators have independent coupling parameters, the network can be presented as connectivity graph \cite{Boccaletti2006}. The Kuramoto model on complex networks is extended from Equation \eqref{eq: Kuramoto1} as:

\begin{equation}
    \frac{d \psi_i}{dt} = \omega_i + \sum_{j=1}^{N}\kappa_{ij}K_{ij} \sin(\psi_j-\psi_i), \quad i = 1,\dots,N,
\end{equation}
where $\lambda_{ij}$ indicates the coupling strength between the nodes $i$ and $j$, and $K$ is the adjacency matrix. The matrix $K$ defines which oscillators are coupled to each other. A three node closed network is illustrated in Figure \ref{fig: Kuramoto} along with its respective adjacency matrix.

\begin{figure}
    \centering
    \begin{subfigure}[c]{0.5\textwidth}
        \centering
        \includegraphics[width=0.6\textwidth]{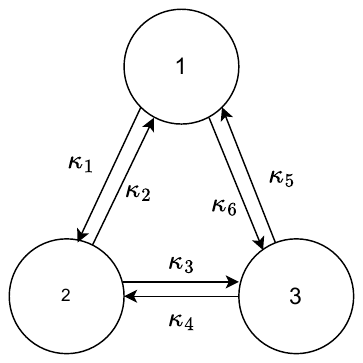}
    \end{subfigure}%
    \begin{subfigure}[c]{0.4\textwidth}
        \centering
        {\normalsize
        $K =
        \begin{pmatrix}
            0 & 1 & 1 \\
            1 & 0 & 1 \\
            1 & 1 & 0
        \end{pmatrix}
        $
        }
    \end{subfigure}
    \caption{Example of a three-node closed network connectivity graph and its adjacency matrix.}
    \label{fig:Kuramoto}
\end{figure}

\paragraph{Critical coupling}

The critical coupling value marks the threshold at which the system transitions from incoherent behavior to collective synchronization. In Kuramoto’s original analysis \cite{Kuramoto_1975}, this threshold is given by
\begin{equation}
K_{\mathrm c}
= \frac{2}{\pi g(0)}
%=\frac{2}{\pi \left(\frac{1}{\sqrt{2\pi}\sigma}\right)}
= 2\sqrt{\frac{2}{\pi}}\sigma
\label{eq: critical}
\end{equation}

Here, $g(0)$ denotes the peak value of the natural frequency distribution \(g(\omega)\). For a Gaussian distribution $g(\omega) \sim \mathcal{N}(\mu, \sigma^2)$, the density attains its maximum at $\omega = \mu$; in the centered case $\mu = 0$, this peak value is $g(0) = 1/(\sqrt{2\pi}\sigma)$, leading directly to \eqref{eq: critical}. Thus, the critical coupling increases linearly with the standard deviation $\sigma$, indicating that broader frequency distributions require stronger coupling to achieve synchronization.

\section{Method}
\label{sec: Method}
We approximate the posterior in Equation~\eqref{eq:posterior} using neural posterior estimation (NPE), through the \textit{BayesFlow} framework for amortized Bayesian inference introduced by Kühmichel et al.\ \cite{Kuhmichel2026bayesflow2}. This avoids the need to evaluate the likelihood function $p(y\mid\theta)$ in order to perform inference. NPE only requires the ability to sample from the joint distribution of the observations and parameters $p(y, \theta)$, which is defined by the simulator $G$ and the prior $p(\theta)$, as:
\begin{equation}
    y = G(\theta, \zeta),\quad \theta\sim p(\theta), \quad \zeta \sim \mathcal{N}\left(0, \sigma^2\right),
\end{equation}
where $\zeta$ is additive Gaussian noise, although other noise models could be used. As Figure \ref{fig: abi illustration} shows, the neural network is trained with data constructed by the simulator that draws the model parameters from a set prior. In the case of a Kuramoto model, we use the prior $p(\kappa) \sim \text{Uniform}(0, u)$, where the upper limit $u$ is determined based on the critical coupling $K_{\mathrm{c}}$. A uniform prior distribution provides a training dataset that equally represents all possible parameter values $\kappa$. 

The most common choice for the inference network is coupling flow, a type of normalizing flow, because it can model complex posterior distributions using simple, invertible transformations that are easy to sample from and efficient to compute, even in high-dimensional settings. \textit{BayesFlow} offers also other options (flow matching, diffusion network) but a short experiment revealed that, in our setting, they did not yield better results than the coupling flow.

Depending on the application, summary statistics can be either handcrafted or determined by a separate summary network. Overall, summary statistics have been shown to be effective for reducing the dimensionality of the original data and for extracting the most informative features from it \cite{Wood2010}. While neural‑network‑derived summary statistics are commonly used in ABI applications, feature engineering of summary statistics can provide significant advantages. In particular, they may accelerate the inference workflow and improve interpretability when designed with domain knowledge in mind \cite{Kuhmichel2026bayesflow2}.

The Kuramoto model is built to simulate and describe the natural synchronization of oscillators. The swarming behavior of these oscillators, as illustrated in Figure \ref{fig: Kuramoto}, is dependent on the order parameter $r$ and the mean phase $\Psi$. For strongly synchronized oscillators $r$ is close to one, for weakly synchronized it is close to zero. The standard deviation of the phase also describes how much the oscillators are spread across the unit circle, since the minimum spread is obtained when all oscillators have identical coordinates. Therefore, we augment the summary statistics with the mean and standard deviation of the coordinates of the oscillators, $(\cos(\psi_i),\sin(\psi_i))$. This results in a total of 6 statistics we can use for inference, listed in Table \ref{tab:summary_statistics}.

\begin{table}[!ht]
    \centering
    \caption{Handcrafted summary statistics.}

    \begin{tabular}{cc}
        \hline
        \multicolumn{2}{c}{Variables} \\
        \hline
        $r$ & $\Psi$  \\
        $\text{mean}(\sin(\psi))$ & $\text{std}(\sin(\psi))$ \\
        $\text{mean}(\cos(\psi))$ & $\text{std}(\cos(\psi))$ \\
        \hline
    \end{tabular}
    \label{tab:summary_statistics}
\end{table}

The issue of selecting suitable summary statistics is pervasive in SBI and related methods, such as approximate Bayesian computation (ABC) \cite{prangle2014abc,nunes2010optimal}.
Learning the summary statistics from data, for example using a neural network, is an attractive option. However, our initial experiments revealed that neural‑network‑derived summary statistics did not yield better results than the handcrafted summary statistics that we obtained through feature engineering based on domain knowledge. Therefore, we chose not to employ a summary network \cite{Kuhmichel2026bayesflow2} whilst training the model. Instead, we use features that directly characterize the swarming behavior of the oscillators. The same features were previously used in MCMC-based studies  \cite{Shah2023kuramoto}. All of these features combined were found to be necessary for successful parameter identification.   

\section{Experiments}
\label{sec: Experiments}
In the following numerical experiments we consider Kuramoto systems with $N$ oscillators, where the goal is to estimate the posterior of the coupling constant $\kappa$ given the simulated data. We create a dataset consisting of $n_t$ simulations from the Kuramoto system for the time period of $\boldsymbol{t} = (t_0, \dots,t_{m})$, $m=1000$. The simulations are divided into 64 equal size batches. From the simulations, we select every $T$-th timestep as an observation. The observation interval, together with dataset size, is especially important for preventing the neural network from overfitting the training data. With a more complex system, more observations and a larger dataset is needed to capture the full nature of the system. For reproducibility, the random seed is set to 41.

The model is trained with data $D_{\text{train}} \in \mathbb{R}^{N \times \frac{m}{T} \times n_t}$. The validation data is created with same setup, with size $n_v$. In both cases the noise is set to $\zeta \sim \mathcal{N}(0,10^{-2})$. The fixed parameters for the simulations and model training are listed in Table \ref{tab: model parameters}. Any other parameters not mentioned here are the default settings by \textit{BayesFlow}.

\begin{table}[!ht]
    \centering
    \caption{Fixed training parameters.}
    \begin{tabular}{c|c|c}
        \hline
         & Simple network & Complex network \\
         \hline
        Oscillators ($N$)  & 100 & 3 \\
        Estimated parameters ($\kappa$) & 1 & 6 \\
        Training samples ($n_t$) & $2^{12}$ & $2^{17}$ \\
        Validation samples ($n_v$) & $2^6$ & $2^7$ \\
        Epochs & 50 & 150 \\
        Batches & 64 & 128 \\
        Initial learning rate & 0.0005 & 0.005 \\
        Dropout & 0.1 & 0.1 \\
        \hline
    \end{tabular}
    \label{tab: model parameters}
\end{table}

In the first simulation model, $\omega$ is drawn at each simulation run from a normal distribution, $\mathcal{N}(1,0.5^2)$. Since the critical coupling value $K_{\mathrm c}$ is dependent on  $g(\omega)$ and $\sigma$ through Equation \eqref{eq: critical}, our theoretical value is $K_{\mathrm c} \approx 0.8$. In practice, the value is slightly higher, but it allows us to set our prior as $p(\kappa) \sim \mathrm{Uniform}(0, 5)$, while ensuring that the system reaches the maximum synchronization state each simulation.

After experimenting with a similar setup for the complex network, the results indicated that the space for all possible combinations of $\omega$ and $\kappa$ is too large to be efficiently approximated by our approach. Therefore each simulation for the complex network is performed with the same value of $\omega$.

\subsection{Results}

Both of the models are evaluated by creating $n_{s} = 300$ simulations for the same time period $\boldsymbol{t}$. Each simulation is then used to draw $n_{d} = 1000$ samples for $\kappa$ from the approximate NPE posterior. 

\begin{figure}[!ht]
    \centering
    \includegraphics[width=0.6\linewidth]{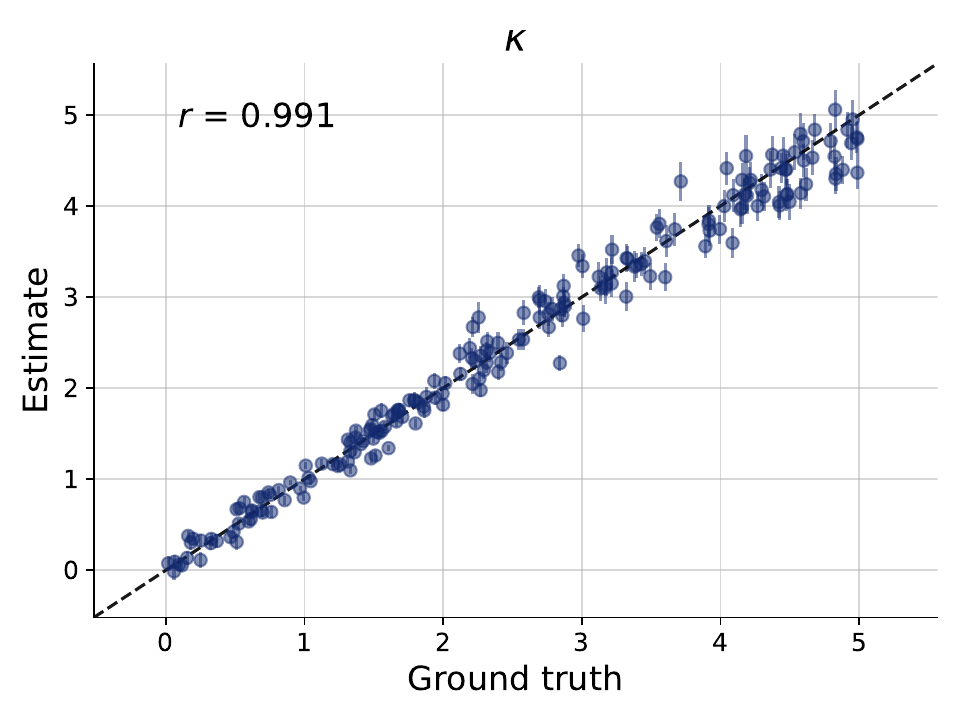}
    \caption{Estimated vs true $\kappa$ values with $n_d$ samples from $n_v$ simulations.}
    \label{fig:estimates}
\end{figure}

The model trained for parameter estimation of the simple first order Kuramoto system successfully recovers the value of the coupling strength $\kappa$ from simulated oscillator trajectories. As illustrated in Figure \ref{fig:estimates}, the predicted values of $\kappa$ align closely with the corresponding ground‑truth parameters across the full range of experimental conditions described in Section \ref{sec: Experiments}. This indicates that the model not only captures the qualitative behavior of the system but also provides quantitatively reliable estimates.

\begin{figure}[!ht]
    \centering
    \includegraphics[width=\linewidth]{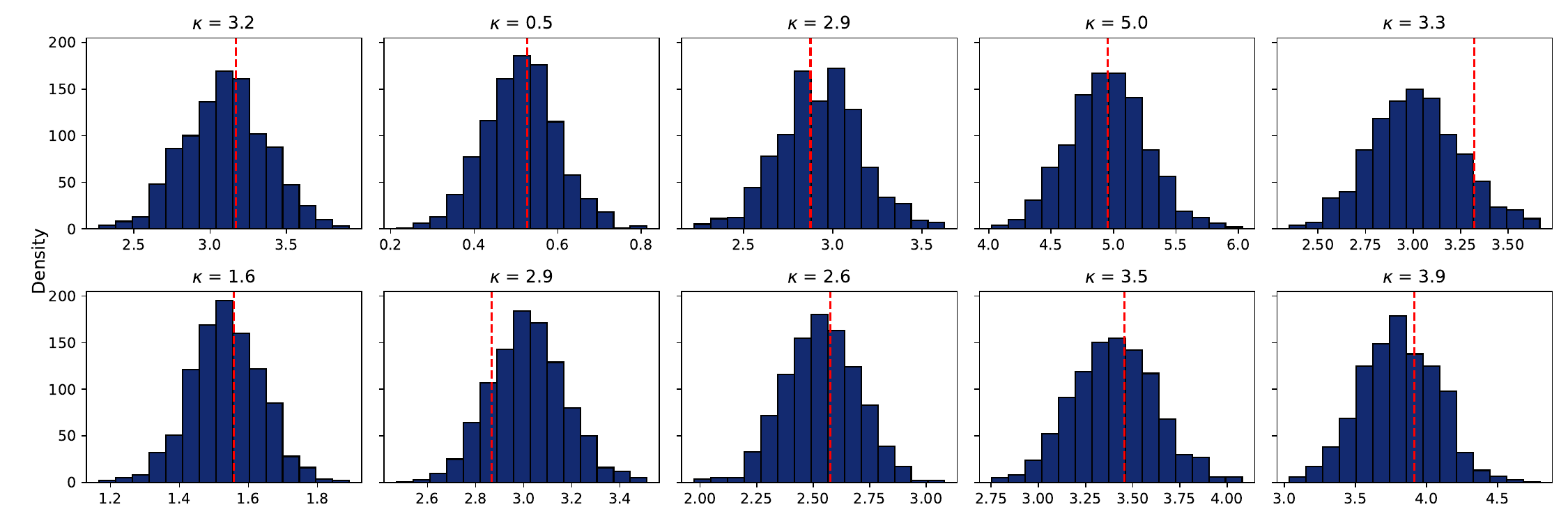}
    \caption{Posterior distribution for $\kappa$ with $n_d$ samples.}
    \label{fig:posterior}
\end{figure}

To further illustrate the model’s predictive behavior, Figure \ref{fig:posterior} presents a more detailed view of the inferred parameter distributions. The histograms show the posterior estimates for 10 different ground‑truth values of $\kappa$, demonstrating how the model expresses uncertainty and how sharply (or broadly) the posterior concentrates around each true parameter. These results highlight both the robustness of the inference procedure and the model’s ability to adapt to variations in the underlying dynamics.

\begin{figure}[!ht]
    \centering
    \begin{subfigure}[t]{0.5\textwidth}
        \centering
        \includegraphics[width=\textwidth]{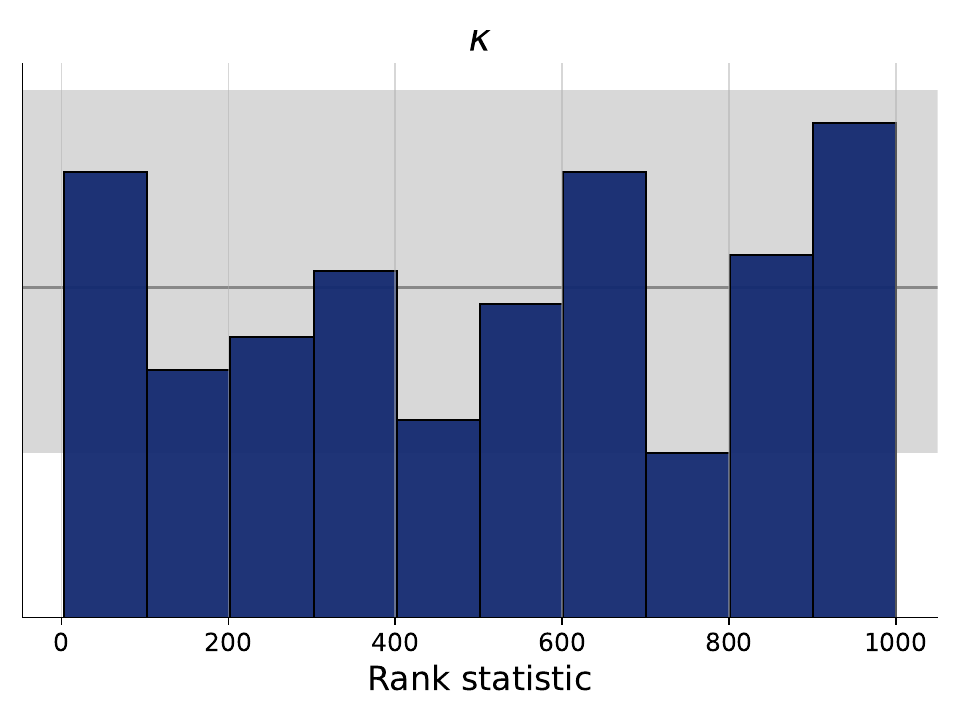}
        \caption{Histogram}
    \end{subfigure}%
    ~ 
    \begin{subfigure}[t]{0.5\textwidth}
        \centering
        \includegraphics[width=\textwidth]{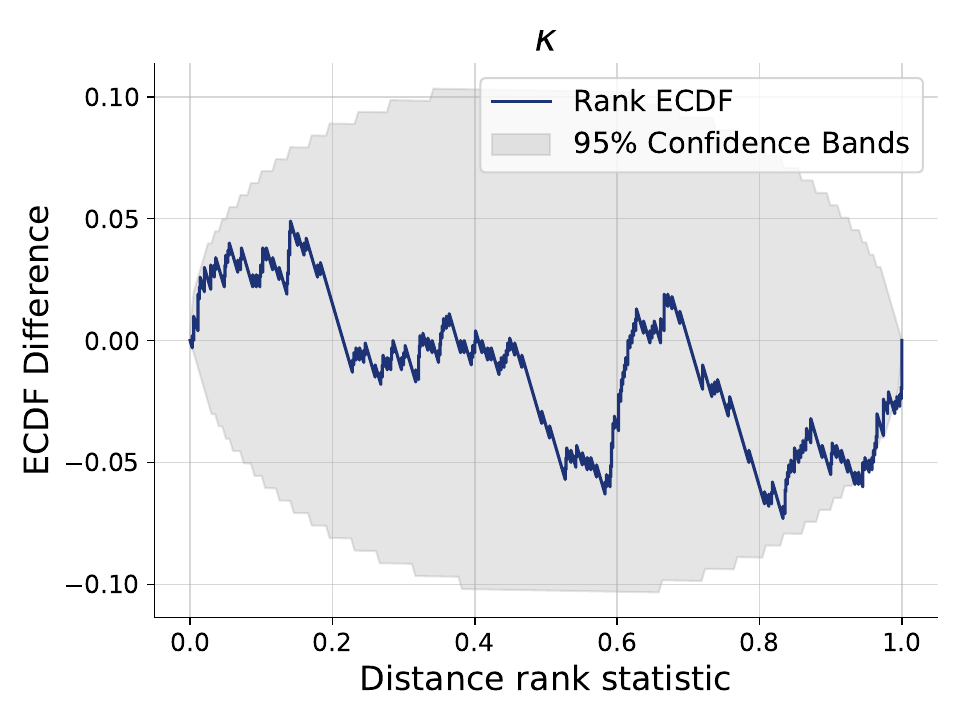}
        \caption{ECDF}
    \end{subfigure}
    \caption{Goodness of fit evaluation.}
    \label{fig: gof}
\end{figure}

The goodness of fit of the models is assessed using two complementary diagnostic measures: the histogram of the probability integral transformation (PIT) values and their empirical cumulative distribution function (ECDF) \cite{Sateilynoja_2022}. The PIT histogram provides a visual check of uniformity, while the ECDF allows deviations from the expected distribution to be examined across the entire range of values. As shown in Figure \ref{fig: gof}, both diagnostics indicate good agreement between the model predictions and the observed data, suggesting that the samples are consistent with being drawn from the same underlying distribution.

\paragraph{Complex network of Kuramoto oscillators}

The model is also tested on an $N=3$ node Kuramoto model with a complex network. The structure of the network is visualized in Figure \ref{fig: Kuramoto} and more detailed structure of $\kappa$ is shown in Equation \eqref{eq: kappas}. Our preliminary experiments showed that the ABI neural network is not capable of reliably estimating the posterior for the $\kappa$ parameters. In the case where $\omega$ is drawn for each simulation separately, this can cause an identifiability problem, where different parameter combinations produce similar outcomes. At the same time, the amount of training data is insufficient to adequately cover the parameter space in order for ABI to accurately learn the mapping function. By limiting the natural frequencies $\omega$ to only one initialization, the results demonstrate that the model captures the essential features of the posterior.

\begin{equation}
            \kappa = 
        \begin{pmatrix}
            0 & \kappa_1 & \kappa_6 \\
            \kappa_2 & 0 & \kappa_3 \\
            \kappa_5 & \kappa_4 & 0
        \end{pmatrix}
        \label{eq: kappas}
\end{equation}

Three evaluation metrics are calculated from the test simulations and summarized in Table \ref{tab:metrics_comparison}: normalized root mean square error (NRMSE), posterior contraction, and calibration error \cite{Kuhmichel2026bayesflow2}. Overall, the metrics for the simple model indicate that the neural network has successfully converged: parameter recovery is accurate, and the posterior distributions are well‑contracted around the ground‑truth values. In contrast, the metrics for the complex network reveal substantially weaker performance. Although the calibration error remains low, the model suffers from poor posterior contraction and a high NRMSE, suggesting that it struggles to accurately infer the underlying parameters in this setting.

\begin{table}[ht]
\centering
\caption{Evaluation of model performances for simple and complex network using normalized root mean square error, posterior contraction and calibration error.}
\label{tab:metrics_comparison}
\begin{tabular}{l|c|cccccc}
\hline
 & Simple network & \multicolumn{6}{c}{Complex network} \\
Metric & $\kappa$& $\kappa_1$ & $\kappa_2$ & $\kappa_3$ & $\kappa_4$ & $\kappa_5$ & $\kappa_6$ \\ \hline
NRMSE & 0.0584 & 0.3139 & 0.3136 & 0.2336 & 0.2206 & 0.2687 & 0.2889 \\
Posterior Contraction & 0.9874 & 0.4469 & 0.2975 & 0.6673 & 0.8335 & 0.5725 & 0.5779 \\
Calibration Error     & 0.0683 & 0.0463 & 0.0181 & 0.0089 & 0.0178 & 0.0423 & 0.0651 \\\hline
\end{tabular}
\end{table}

\begin{figure}
    \centering
    \includegraphics[width=\linewidth]{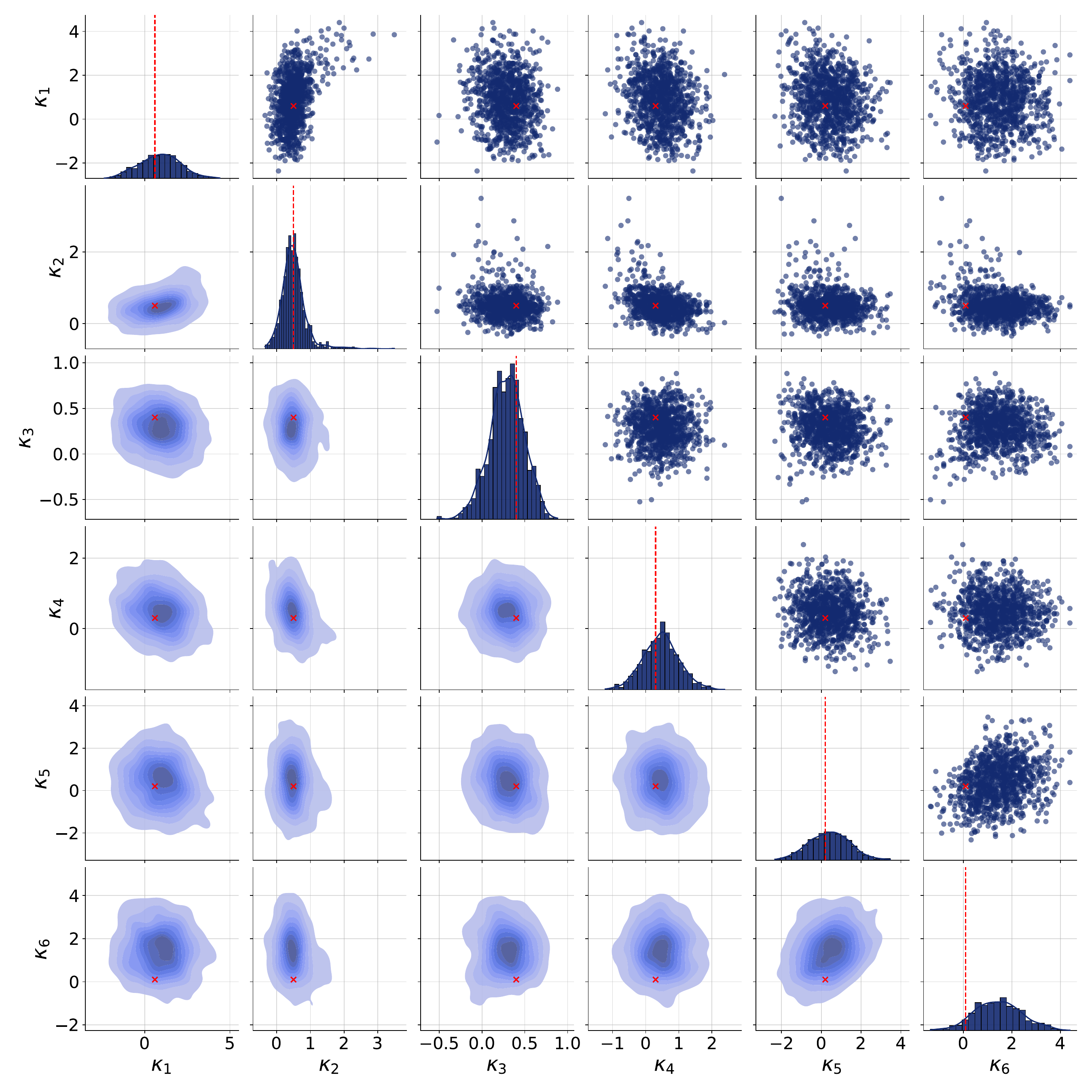}
    \caption{Posterior estimates for the complex network with true parameters $\kappa_1 = 0.6$, $\kappa_2 = 0.5$, $\kappa_3=0.4$, $\kappa_4=0.3$, $\kappa_5=0.2$ and $\kappa_6=0.1$. The true parameter values are marked with red dashed lines or circular red dots.}
    \label{fig: complex posterior}
\end{figure}

\paragraph{Comparison to MCMC}

The ABI's capability to capture the posterior $p(\kappa \mid y)$ is compared with the method presented by \cite{shah2023fods}, where the likelihood is constructed from an empirical cumulative distribution function (ECDF) of the same features as those used above. Since the Kuramoto simulations are randomized, a  standard likelihood based on the residuals between data and simulation values is not available. Instead, we characterize the statistical distribution of the features by their ECDF vectors. The vectors provide a stochastic Gaussian likelihood, see \cite{haario2015generalized, Springer2019} where the approach was first applied to chaotic dynamical systems.  The results  we use here for synchronization  are given in PhD thesis  \cite{Shah2023kuramoto}. The Gaussian likelihood of the ECDF vectors was able to capture the posterior for a three node closed network and identify all of the parameters.

By running ABI with a similar setting, we are able to identify the main differences between our proposed approach and the ECDF posterior. We create a dataset with known parameters $\kappa_1 = 0.6$, $\kappa_2 = 0.5$, $\kappa_3=0.4$, $\kappa_4=0.3$, $\kappa_5=0.2$ and $\kappa_6=0.1$, and use the trained model to give us estimated values. As seen in Figure \ref{fig: complex posterior}, the ABI is not able to identify all of the parameters with the same level of accuracy as the ECDF approach. The posterior distribution is too wide and the correlation found by Shah in \cite{Shah2023kuramoto} is not visible.

\section{Conclusions and future work}
\label{sec: Conclusions}

The introduced ABI model successfully captures the posterior for the simple first order Kuramoto system. As the approach requires training the neural network, comparison to ECDF posterior estimation revealed that while the approach is able to capture the posterior for single parameter model, the computational for training the neural network is high. It should be noted that whilst the ECDF is constructed for a single sample, ABI trains the neural network with parameter combinations from the prior.

The degradation in posterior contraction in the complex network case suggests a combination of structural non-identifiability and amortization-induced smoothing. In particular, the permutation symmetry of node labels and the confounding between coupling strengths and natural frequencies reduce the effective information content of the summary statistics. As a result, the amortized posterior estimator exhibits increased variance and limited parameter recovery despite acceptable calibration. Although the results are promising, they primarily highlight the need for further exploration of alternative approaches within the ABI framework. Also, it is worth mentioning that even more careful selection of summary variables, or combining it with the summary network, might yield better results.

Since the proposed ABI model does not successfully recover the posterior for the complex Kuramoto network, one natural direction is to extend the method—for example through preconditioned neural posterior estimation (NPE) \cite{wang2024preconditioned}. Neural network based inference methods are generally prone to model misspecification, and several strategies have been proposed to mitigate this issue. These include model comparison through self‑consistency checks \cite{kucharsky2025improving} as well as approaches based on generalized Bayesian inference \cite{Sun2026amortized, Gao2023generalized}. Together, these methods provide potential routes for improving posterior accuracy in more demanding dynamical systems.

\section{Acknowledgments}
This work was supported by the Finnish Ministry of Education and Culture’s Pilot for Doctoral Programmes (Pilot project Mathematics of Sensing, Imaging and Modelling) and Research Council of Finland (Flagship of Advanced Mathematics for Sensing and Imaging and Modeling grant 359183). %, Centre of Excellence of Inverse Modelling and Imaging grant 353095). 
The research of Jana de Wiljes has been partially funded by the Deutsche Forschungsgemeinschaft (DFG)- Project-ID 318763901 - SFB1294. Furthermore, this project has received funding from the European Union under the Horizon Europe Research \& Innovation Programme (Grant Agreement no. No 101188131 UrbanAIR). Views and opinions expressed are however those of author(s) only and do not necessarily reflect those of the European Union. Neither the European Union nor the granting authority can be held responsible for them.

\addcontentsline{toc}{section}{References}
\bibliographystyle{model1-num-names}
\bibliography{refs.bib}
\end{document}